# An algorithm of finding rules for a class of cellular automata

## Lei Kou

Institute of Oceanographic Instrumentation,

Qilu University of Technology (Shandong Academy of Sciences),

Qingdao, China

Email: koulei1991@hotmail.com or koulei1991@qlu.edu.cn

## Fangfang Zhang* and Luobing Chen

School of Electrical Engineering and Automation,

Qilu University of Technology (Shandong Academy of Sciences),

Jinan, China

Email: zhff4u@qlu.edu.cn

Email: a18730280565@163.com

## Wende Ke

Department of Mechanical and Energy Engineering,

Southern University of Science and Technology,

Shenzhen, China

Email: kewd@sustech.edu.cn

## Quande Yuan

School of Computer Technology and Engineering,

Changchun Institute of Technology,

Changchun, China

Email: yuanqd@airlab.ac.cn

## Junhe Wan and Zhen Wang

Institute of Oceanographic Instrumentation,

Qilu University of Technology (Shandong Academy of Sciences),

Qingdao, China

Email: wan_junhe@qlu.edu.cn

Email: wangzhen_82@126.com

**Abstract:** Cellular automata (CA) is an important modelling paradigm for complex systems. In the design of cellular automata, the most difficult task is to find the transformation rules that describe the temporal evolution or pattern of a modelled system. A CA with weights(CAW) yields transition rules algorithm is proposed in this paper, which have ample physical meanings and extend the category of CA. Firstly, the weights are increased to connect the updated cell and its neighbours, and the output of each cell depends on the states of cells in the neighbourhood and their respective weights. Secondly, the error correction algorithm is adopted to find correct transition rules by adjusting weights. When the error is zero, the required transition rules with correct weights will be found to describe the fixed configuration. The CAW with the correct rules will relax to the fixed configuration regardless of the initial states. Finally, the mathematical analysis and simulation are carried out with one-dimensional CAW, and the results show that the proposed algorithm has the ability to find correct transition rules as the error converges exponentially.

**Biographical notes:** Lei Kou received the M.Sc. degree in Computer Science and Technology from Northeast Electric Power University in 2017, and the Ph.D. degree in Electrical Engineering from Northeast Electric Power University in 2020. He is a research associate at Institute of Oceanographic Instrumentation, Qilu University of Technology (Shandong Academy of Sciences). His current research interests include artificial intelligence, machine learning.

Fangfang Zhang received the Ph.D. degree from Shandong University. She is an Associate Professor with the School of Electrical Engineering and Automation, Qilu University of Technology (Shandong Academy of Sciences). Her research interests include the complex chaotic systems, and chaotic neural networks.

Luobing Chen is currently studying for the M.Sc. degree in School of Electrical Engineering and Automation, Qilu University of Technology (Shandong Academy of Sciences), Jinan, China. Her research interests include chaotic neural networks and convolutional neural networks.

Wende Ke received the Ph.D. degree in Computer Science from Harbin Institute of Technology. He is a Professor of Southern University of Science and Technology, Shenzhen, China. His current research interests include artificial intelligence and robotics.

Quande Yuan received the Ph.D. degree in Computer Science from Harbin Institute of Technology. He is an associate professor at Changchun Institute of Technology. His current research interests include multi robot system, machine learning.

Junhe Wan received her Ph.D. degree in Intelligent Information and Communication System from Ocean University of China in 2020. She is a research associate at Institute of Oceanographic Instrumentation, Qilu University of Technology (Shandong Academy of Sciences). Her research interest includes artificial intelligence, machine learning.

Zhen Wang received the Ph.D. degree from Harbin Institute of Technology in 2011. He is an associate professor at Institute of Oceanographic Instrumentation, Qilu University of Technology (Shandong Academy of Sciences). His current research interests include artificial intelligence, machine learning, intelligent monitoring.

# 1   Introduction

Cellular Automaton (CA) is one of the most popular methods for studying complex systems. It was first introduced by Von Neumann as a tool for studying biological systems [1]-[2] and since then, it has been utilized extensively to study complex systems. The basic idea is quite simple: consider a collection of cells with defined finite states. The evolution of the system is modelled by updating the state of each cell based on pre-described rules (i.e. transition rules) which take into account the states of its immediate neighbours. Cellular automata has been successfully applied in many fields.

In terms of forest fire, Gheorghe et al. [3] developed a forest fire model using the cellular automata algorithm framework. The model is essentially a learning and training tool. It aims to explain various physical factors and variables in the process of fire ignition, spread, and extinguishing in an interactive way. Zhou et al. [4] proposed a three-dimensional cell space forest fire spread model based on a multi-objective genetic algorithm. The experimental results of the proposed model are highly similar to the actual forest fire spread process, and the model can simulate the dynamic process of forest fire spread in detail.

In terms of image processing, Piao et al. [5] proposed a light field significance detection method based on the model of deep induced cellular automata. The proposed method is robust to a series of challenging scenes, such as cluttered background areas, background areas with similar appearance and depth to salient objects, and so on. Trevisi et al. [6] proposed a hardware friendly pseudo-random ternary measurement matrix based on cellular automata (ECA). The model is used for compressed sensing acquisition of natural images. Sanchez-Macian et al. [7] proposed quantum-dot cellular automata (QCA) as a candidate circuit to replace traditional CMOS circuits. Applying QCA approximate adder to image processing can reduce the average normalized average error distance.

In terms of transportation, Zhang et al. [8] proposed an improved single lane cellular automata traffic flow model. The high-speed car following rate of the model is in good agreement with the measured traffic results. Enayatollahi et al. [9] developed a 2D cellular automata model for regional navigation, which covers the area where the aircraft enters the arrival route until the initial approach waypoint. The model aims to help minimize the total delay time of air traffic. Lv et al. [10] established a mesoscopic cellular automata model for large-scale evacuation scenarios. Based on the simulation results, the possible problems in the actual evacuation process can be found, and the corresponding improvement guidelines and suggestions can be put forward.

In wastewater treatment, Zhang et al. [11] established the cellular automata model of sequencing batch reactor according to the biological mechanism of the activated sludge process. The model reproduces the complex evolution process of microorganisms in the reaction process, and describes the removal of biochemical oxygen demand and the growth of activated sludge. Qiao et al. [12] proposed a simulation method of a three-dimensional cellular automata model for the treatment process of activated sludge process. This method can effectively reflect the treatment process of the activated sludge process.

In terms of biological systems, Tsompanas et al. [13] proposed a physically inspired network design model based on cellular automata. The proposed CA model can be used as a virtual, accessible, and bionic laboratory simulator, so as to save a lot of time for biological experiments. Shimada et al. [14] proposed a new multi-ventricular neuron model (a new multi-ventricular dendritic cell body model). The dynamic characteristics of the model are described by coupled asynchronous cellular automata. The results show that the model can reproduce the typical dendritic phenomena observed in biological neurons and multi ventricular neuron models. Matsubara et al. [15] proposed a neuron model based on asynchronous cellular automata, which is described as a special kind of cellular automata. The model contributes to the development of engineering and clinical applications. Based on cellular automata and fast marching method, Sallemi et al. [16] developed a new model to simulate tumor growth and used it to estimate the evolution of brain tumors during this period. This study can help radiologists make a more accurate and objective diagnosis. Takeda et al. [17] proposed a cochlear model based on nonlinear dynamics of asynchronous cellular automata. Theoretical divergence analysis shows that the model can simulate the nonlinear vector field of differential equation cochlear model and reproduce the frequency tuning curves of many organisms. Zhang et al. [18] established an ECG cellular automata model including atrial muscle, ventricular muscle, atrioventricular cavity, and ventricular septum. The results show that the model can accurately clarify the relationship between ECG and cardiomyocyte electrical activity.

However, the most difficult task in the design of cellular automata is to find a transition rule that will describe the temporal evolution of a modelled system, i.e. which leads to desired global response of the system. As the number of possible rules dramatically increases with the number of states and the size of the neighbourhood, it is usually non-trivial to find correct transition rules describing the system being modelled.

Wolfram [19] proposed four classes of cellular automata behaviour: (1) almost all configurations relax to a fixed configuration, (2) almost all configurations relax to either a fixed point or a periodic cycle according to the initial configuration, (3) almost all configurations relax to chaotic behaviour, (4) sometimes initial configurations produce complex structures that might be persisting or long-living.

In this paper, as for the first class of CAs which relax to a fixed configuration, we present a method to design transition rules based on optimization algorithms [20-22]. First of all, we add connection weights between the updated cell and other cells in the neighbourhood, and the output of the updated cell depends on states of the cells in the neighbourhood and their respective weights. We call it CA with weights(CAW), which extend the category of CA. Then we calculate the error criterion function, which is produced by real configuration and desired fixed configuration, and adopt error correction algorithm to modify corresponding weights until we find transition rules with correct weights. The CA with the correct rules will relax to the fixed configuration regardless of the initial configuration.

The paper is organized as follows: in Section 2, the evolution of CA with weights is introduced in details and its characteristics are compared with neural networks and other models. In Section 3, we describe the algorithm of finding rules and prove its convergence with an exponential speed. In Section 4, we make simulations and verity the correctness and ability of CAW. Finally, some conclusions and applications of CAW are discussed in Section 5.

## 2     The evolution of CA with weights and its characteristics

### 2.1  The structure of CAW

Cellular automata (CA) is a method used to simulate local rules and local connections. A typical cellular automaton is defined on a grid. The grid at each point represents a cell with a finite state. The transition rules apply to each cell and are performed simultaneously. Here is a stricter definition of cellular automaton.

**Definition1:** Generally, cellular automaton satisfies:

(1) The regular cells cover a part of d dimensional space;

(2) $x_k(i_1, i_2, ..., i_d)$ represent the states of $Cell(i_1, i_2, ..., i_d)$ $x_k(i_1, i_2, ..., i_d) \in \{0, 1\}$, $r$ indicates the radius of the neighbourhood. The type of neighbourhood depends on the special system.

(3) Transition functions f specify the evolution of $x_k(i_1, i_2, ..., i_d)$ are as follows,

$$x_{k+1}(i_1, i_2, ..., i_d) = f(x_k(i_1 - r, i_2, ..., i_d), ..., x_k(i_1, i_2, ..., i_d + r)) \quad (1)$$

where the output of $f()$ either is 0 or 1.

Based on CA, we add connection weights between the updated cell and other cells in the neighbourhood, and extend the output of the updated cell from the set $\{0, 1\}$ to the real filed. We call it CA with weights (CAW).

The evolution diagram of one-dimensional CAW (with radius equal to one) is depicted in Figure 1(a). The dotted lattices in the boundary indicate virtual cells. Boundary conditions can be periodic, fixed or reflecting among others. The state of $Cell(i)$ at step k is $x_k(i)$. There are three cells in the neighbourhood and their weights are $w(i, u)$, $u = i-1, i, i+1$. Each cell, the input of which is related to the states of the cells in the neighbourhood and their weights, has a transition function and output $x_{k+1}(i)$. It means

$$x_{k+1}(i) = f(x_k(i-1), x_k(i), x_k(i+1))$$
$$= f(x_k(i-1)w(i, i-1) + x_k(i)w(i, i) + x_k(i+1)w(i, i+1)).$$

Figure 2 shows the evolution of two-dimensional CAW with Von Neumann neighbours (with radius equal to one), in which only the nearest neighbors of the updated cell are involved (i.e. four plus one in total). The input of transition function of $Cell(i, j)$ depends on the states of these five cells at step k and their weights. The $r$-neighborhood of a cell $Cell(i, j)$ with Von Neumann neighbours is defined as follows,

$$N_r(i, j) = \{Cell(u, v) \big| |i - u| + |j - v| \le r\} \quad (1)$$

The $r$-neighborhood of a cell $Cell(i, j)$ with Moore neighbours is defined as follows,

$$N_r(i, j) = \{Cell(u, v) \big| |i - u| \le r, |j - v| \le r\} \quad (2)$$

**Figure 1** The evolution diagram of one-dimensional CAW ($r$=1) and discrete Hopfield neural network

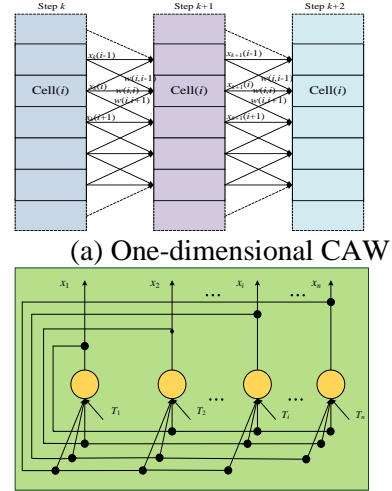

(a) One-dimensional CAW

(b) Discrete Hopfield neural network

**Figure 2**  The evolution diagram of two-dimensional CAW with Von Neumann neighbourhood ($r$=1)

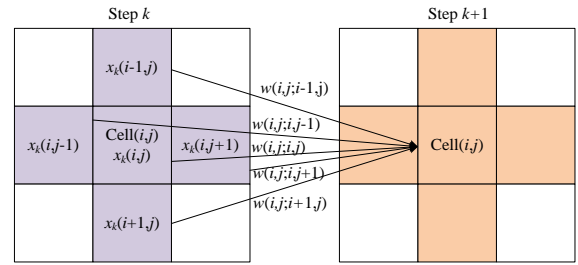

It is easy to understand for one-dimensional CAW, $R=r$. As for two-dimensional CAW with Von Neumann neighbourhood, $R=(2r-1)^2/2+2$ (we take its integer part, hereinafter the same); for two-dimensional CAW with Moore Neumann neighbourhood, $R=(2r+1)^2/2$; for $d$-dimensional CAW with Von Neumann neighbourhood $R=(2r-1)^d/2+d$; for $d$-dimensional CAW with Moore Neumann neighbourhood $R=(2r+1)^d/2$.

Here is a more strict definition of cellular automaton with weights.

**Definition 2**: Cellular Automaton with Weights

Generally, cellular automaton with weights satisfies:

(1) The regular cells cover a part of $d$ dimensional space;

(2) $x_k(i_1, i_2, \cdots, i_d)$ represent the states of $Cell(i_1, i_2, \cdots, i_d)$.

$w(i_1, i_2, ..., i_d; u_1, u_2, ... u_d), u_1 = i_1 - r, ..., i_1, i_1 + r, u_d = i_d - r, ..., i_d, i_d + r$ represents the weight of each cell in the neighbourhood connected with the updated cell. $x_k(i_1, i_2, ..., i_d) \in R$, $w_k(i_1, i_2, ..., i_d; u_1, u_2, ...u_d) \in R$, $r$ indicates the radius of the neighbourhood. The type of neighbourhood depends on the special system.

(3) The following transition function $f$ specifies the evolution of $x_k(i_1, i_2, ..., i_d)$

$$x_{k+1}(i_1, i_2, ..., i_d) = f(x_k(i_1 - r, i_2, ..., i_d), ..., \qquad (3)$$
$$x_k(i_1, i_2, ..., i_d + r), w(i_1, i_2, ..., i_d, u_1, u_2, ..., u_d))$$

where $f$ can be any expression, including logical expressions and functions which have derivatives. The transition functions of all cells may be the same or different.

It can be seen that classical cellular automaton is the CAW with discrete states, 0 or 1 weights. Therefore, CAW extends the category of CA, and weights has an important physical meaning. They show degree of effects between the cells in the neighbourhood. This is similar to real system such as the society. If we consider every person as one cell, and consider his friends and family as its neighbours, the degree of effects of the neighbours impressed on his growth can be indicated by these weights. It is noted that $w(i, u)$ is different from $w(u, i)$, because the impact of $Cell(i)$ impressed on $Cell(u)$ is not equal to the impact of $Cell(u)$ impressed on $Cell(i)$. In fact, the ability of every cell to learn from its neighbours is different, which indicate the diversity of individuals.

In practice, the weights can be identified by the mechanism of modelled system, such as cellular neural networks, hopfield neural networks, which will be explained in Section 2.2. If the mechanism of modelled system is not known, an algorithm of finding weights is given in Section 3.1 for the system relax to a fixed configuration.

### 2.2 The Characteristics of CAW

CAW has characteristics such as discrete, synchronization and localized interactions as well as classical CA, which is a special CAW in essence.

As there exist weights in neural networks, it is necessary for us to compare CAW with neural networks. The differences between them are as follows:

(1) CAW is a spatially and temporally discrete mathematical model, while some neural networks are continuous and some are discrete such as discrete hopfield neural network.

(2) Each cell of CAW only depends on the cells in the neighbourhood with neighbourhood with a variety of the spatial distribution and their respective weights, not all cells; while the inputs of nodes in neural networks are related to all neurons in the forward layer and their respective weights;

(3) CAW is a model of physical system or process and evolves with time, without $f$; while neural networks learn to approximate nonlinear functions or relax to stable states throughout training weights;

Note that, one dimensional CAW is similar to discrete Hopfield neural network, but there are some differences in excitation function, neuronal mechanism and weight. The diagrams of one-dimensional CAW and discrete Hopfield neural network are shown in Figure 1.

Furthermore, the CAW is similar to cellular neural network (CNN). CNN is a large-scale nonlinear analog circuit which processes signals in real time and made of a massive aggregate of regularly spaced circuit clones, called cells, which communicate with each other directly only through its nearest neighbours. Each cell is made of a linear capacitor, a nonlinear voltage-controlled current source, and a few resistive linear circuit elements. The states equation of CNN can be denoted as

$$\frac{dx(i, j)}{dt} = -x(i, j) + \sum_{Cell(u,v) \in N_r(i,j)} A(i, j; u, v) y(x(u, v))$$
$$+ \sum_{Cell(u,v) \in N_r(i,j)} B(i, j; u, v) u(u, v) + I \qquad (4)$$

where $x(i, j)$ is the state of $Cell(i, j)$, $y(x(u, v))$ is the output of the cell in the neighborhood, $u(u, v)$ is the output of the cell in the neighborhood. The dynamics of a cellular neural network has both output feedback operator $A(i, j; u, v)$ and input control operator $B(i, j; u, v)$ which indicate the interactions among the cells in the neighborhood. It means $w(i, j; u, v)$ in CNN is identified by a vector of $\{A(i, j; u, v), B(i, j; u, v)\}$. Comparing (3) and (4), we can see a remarkable similarity between them. The main difference is that for CAW, the time is discrete and the dynamic function can be any expression including a logic function and real function of the previous states of the neighbour cells, whereas for cellular neural networks, the time is continuous and the dynamic function is a nonlinear real function of the previous states of the neighbour cells. Therefore, to some extent, cellular neural network is a continuous CAW applied to analog circuit.

The comparisons of four mathematical models are summarized in Table 1.

## 3 The algorithm of finding transition rules and its analysis

### 3.1 The algorithm of finding transition rules based on CAW

Here we present an algorithm of finding transition rules which is only applied for the first class CAs. The basic idea is to find correct weights according to delta error correction algorithm based on the structure of CAW.

For simplicity, we take a one-dimensional CAW with radius $r$ and $N$ cells for example. Its diagram is shown in Figure 1(a). The state of $Cell(i)$ at step $k$ is $x_k(i)$, and the weights between $2r+1$ cells in the neighbourhood are $w_k(i,u), u=i-r, \mathrm{L}, i, \mathrm{L}, i+r$. The given state of $Cell(i)$ is $d(i)$ and its rule function is $f$. The process of finding correct weights is as follows.

Inspired by neurons, we assume the cell, the input of which is the sum of products that states of the cells in the neighbourhood multiply corresponding weights, has a structure and function like that of neuron. Assume the threshold value of all cells are zero, so the state of $Cell(i)$ at next step is

$$x_{k+1}(i) = f(s_k(i)) = f(\sum_{u=i-r}^{i+r} x_k(u)w_k(i,u)) \quad (5)$$

where $x_k(u)$ is state of the cell which belongs to the neighbourhood of $Cell(i)$ and its weight is $w_k(i,u)$ at step $k$.

The whole error criterion function is defined as

$$E_{k+1} = \sum_{i=1}^{N} e_{k+1}(i) = \frac{1}{2}\sum_{i=1}^{N}[d(i)-x_{k+1}(i)]^2 \quad (6)$$

From gradient descent method, we get

$$\begin{aligned}
\Delta w_k(i,u) &= -\frac{\partial e_{k+1}(i)}{\partial w_k(i,u)} \\
&= -\frac{\partial e_{k+1}(i)}{\partial x_{k+1}(i)}\frac{\partial x_{k+1}(i)}{\partial s_k(i)}\frac{\partial s_k(i)}{\partial w_k(i,u)} \quad (7) \\
&= [d(i)-x_{k+1}(i)]f'(s_k(i))x_k(u)
\end{aligned}$$

Therefore, the weights of $Cell(i)$ at step $k+1$ are

**Table 1.** Comparisons of four mathematical models

| Model | Cellular automata with weights | Classical cellular automata | Neural networks | Cellular neural network |
|---|---|---|---|---|
| Time | discrete | discrete | Discrete or continuous | continuous |
| Space | discrete | discrete | discrete | discrete |
| State | Real | discrete | real | real |
| weights | Real | 1 or 0 | real | real |
| dynamics | nonlinear | nonlinear | nonlinear | nonlinear |

For $Cell(i)$,

$$\begin{aligned}
e_{k+1}(i) &= \frac{1}{2}[d(i)-x_{k+1}(i)]^2 \\
&= \frac{1}{2}[d(i)-f(\sum_{u=i-R}^{i+R} x_k(u)w_k(i,u))]^2, i=1,2,\cdots,m
\end{aligned} \quad (9)$$

We assume there is a beeline between $e_{k+1}(i)$ and $e_k(i)$, which does not affect the convergence of CAW. Therefore, we consider the error criterion function as a

$$\begin{aligned}
w_{k+1}(i,u) &= w_k(i,u) + \eta\Delta w_k(i,u) \\
&= w_k(i,u) + \eta[d(i)-x_{k+1}(i)]f'(s_k(i))x_k(u)
\end{aligned} \quad (8)$$

where $u=i-r,...,i,...,i+r$, $\eta$ is the learning factor, and it is a positive constant.

After finite steps, when the error is zero, the weights won't change. We will find the correct weights $w_c(i,u)$. This algorithm is summarized as follows:

Step1: Set $k=1$, and initialize $r$, $N$, $w_k(i,u)$, $\eta$ and $\varepsilon$ (an arbitrary small number);

Step2: Input $x_k(i)$ to CAW and generate $x_{k+1}(i)$ according to (5);

Step3: Calculate $E_{k+1}$ according to (6);

Step4: Calculate $w_{k+1}(i,u)$ according to (8);

Step5: if $E_{k+1} \le \varepsilon$, then output $w_c(i,u)=w_{k+1}(i,u)$; else $k=k+1$, and return step2.

The rule with correct weights is the transition rules we required. Regardless of initial states, CAW with the correct rules will relax to the fixed configuration after several iterations.

Similarly, as for two-dimensional CAW with $N\times M$ cells, we can adopt this algorithm to deduce $w_c(i,j;u,v)$, too. This algorithm also can be applied to higher dimensional CAW.

### 3.2 The convergence of CAW

For convenience, no matter how many dimensions the CAW has, we denote all cells in one dimensional space as $Cell(i), i=1,2,\cdots,m$ where $m$ is the number of cells in the whole space.

continuous function. $e(i)$ is regarded as a function of variable $w(i,u)$, and it is written as $e_i(w)$. Therefore,

$$\begin{aligned}
e_k(i) &= e_i(w_{k-1}) = \frac{1}{2}[d(i)-x_k(i)]^2 \\
&= \frac{1}{2}[d(i)-f(\sum_{u=i-R}^{i+R} x_{k-1}(u))w_{k-1}(i,u))]^2, i=1,2,\cdots,m
\end{aligned} \quad (10)$$

The Taylor series of $e_i(w_k)$ at $w_{k-1}$ is

$$
\begin{aligned}
e_i(w_k) &= e_i(w_{k-1} + \eta \Delta w_{k-1}) \\
&= e_i(w_{k-1}) + \eta \nabla e_i(w)^T \Big|_{w = w_{k-1}} \Delta w_{k-1} \\
&\quad + \frac{1}{2}\eta^2 (\Delta w_{k-1})^T \nabla^2 e_i(w)^T \Big|_{w = \varepsilon_{k-1}} (\Delta w_{k-1})
\end{aligned} \tag{11}
$$

where $\nabla e_i(w) = \begin{bmatrix} \dfrac{\partial e_i}{\partial w(i,i-R)} & \dfrac{\partial e_i}{\partial w(i,i-R+1)} & \cdots & \dfrac{\partial e_i}{\partial w(i,i+R)} \end{bmatrix}^T$ (12)

$$
\nabla^2 e_i(w) = \begin{bmatrix}
\dfrac{\partial^2 e_i}{\partial w(i,i-R)^2} & \dfrac{\partial^2 e_i}{\partial w(i,i-R)\partial w(i,i-R+1)} & \cdots & \dfrac{\partial^2 e_i}{\partial w(i,i-R)\partial w(i,i+R)} \\
\dfrac{\partial^2 e_i}{\partial w(i,i-R+1)\partial w(i,i-R)} & \dfrac{\partial^2 e_i}{\partial w(i,i-R+1)^2} & \cdots & \dfrac{\partial^2 e_i}{\partial w(i,i-R+1)\partial w(i,i+R)} \\
\vdots & \vdots & & \vdots \\
\dfrac{\partial^2 e_i}{\partial w(i,i+R)\partial w(i,i-R)} & \dfrac{\partial^2 e_i}{\partial w(i,i+R)\partial w(i,i-R+1)} & \cdots & \dfrac{\partial^2 e_i}{\partial w(i,i+R)^2}
\end{bmatrix} \tag{13}
$$

$$
\varepsilon_{k-1} = \theta w_k + (1-\theta) w_{k-1}, \ 0 < \theta < 1 \tag{14}
$$

From the algorithm in section 3.1, we have

$$
\nabla e_i(w)\Big|_{w=w_{k-1}} = \frac{\partial e_i}{\partial w}\Big|_{w=w_{k-1}} = -\Delta w_{k-1} \tag{15}
$$

Substituting (15) into (11), we get

$$
\begin{aligned}
e_i(w_k) - e_i(w_{k-1}) &= -\eta(\Delta w_{k-1})^T \Delta w_{k-1} \\
&\quad + \frac{1}{2}\eta^2 (\Delta w_{k-1})^T \nabla^2 e_i(w)^T \Big|_{w=\varepsilon_{k-1}} (\Delta w_{k-1})
\end{aligned} \tag{16}
$$

Because

$$
\begin{aligned}
(\Delta w_{k-1})^T \Delta w_{k-1} &= \sum_{u=1}^{q}(-\frac{\partial e_i}{\partial w_{k-1}(i,u)})^2 \\
&= (-\frac{\partial e_i}{\partial w_{k-1}(i,1)})^2 + (-\frac{\partial e_i}{\partial w_{k-1}(i,2)})^2 \\
&\quad + \cdots + (-\frac{\partial e_i}{\partial w_{k-1}(i,q)})^2 \geq 0
\end{aligned} \tag{17}
$$

the result of (16) depends on the second term on the right. We discuss it in two conditions.

(1) If $\nabla^2 e_i(w)^T \Big|_{w=\varepsilon_{k-1}}$ is negative semidefinite and has a maximum eigenvalue $\lambda_{\max}$ ($\lambda_{\max} \leq 0$), then (16) becomes

$$
\begin{aligned}
e_i(w_k) - e_i(w_{k-1}) &\leq -\eta(\Delta w_{k-1})^T \Delta w_{k-1} + \frac{1}{2}\eta^2 \lambda_{\max}(\Delta w_{k-1})^T (\Delta w_{k-1}) \\
&= -\eta(\Delta w_{k-1})^T (\Delta w_{k-1})(1-\frac{1}{2}\eta\lambda_{\max}) \leq 0
\end{aligned}
$$
(18)

(2) If $\nabla^2 e_i(w)^T \Big|_{w=\varepsilon_{k-1}}$ is not negative semidefinite, we assume

$$
0 \leq C = \left\| \nabla^2 e_i(w) \right\|_2 = \sqrt{\sum_{u=1}^{q}\sum_{v=1}^{q}(\frac{\partial^2 e}{\partial w(i,u)\partial w(i,v)})^2} \leq \frac{2}{\eta} \tag{19}
$$

Then (16) becomes

$$
\begin{aligned}
e_i(w_k) - e_i(w_{k-1}) &\leq -\eta(\Delta w_{k-1})^T \Delta w_{k-1} + \frac{1}{2}\eta^2 C(\Delta w_{k-1})^T (\Delta w_{k-1}) \\
&= -\eta(\Delta w_{k-1})^T (\Delta w_{k-1})(1-\frac{1}{2}\eta C) \leq 0
\end{aligned}
$$
(20)

Compare (18) and (20), we write their unified form as

$$
e_i(w_k) - e_i(w_{k-1}) \leq -\eta(\Delta w_{k-1})^T (\Delta w_{k-1})(1-\frac{1}{2}\eta z) \tag{21}
$$

If $\nabla^2 e_i(w)^T \Big|_{w=\varepsilon_{k-1}}$ is negative semidefinite, $z$ denotes its maximum eigenvalue; else $z$ denotes its norm.

From (7), we know

$$
(-\frac{\partial e_i}{\partial w_{k-1}(i,u)})^2 = [d(i) - x_k(i)]^2 f'^2(s_{k-1}(i) x_{k-1}^2(u), \tag{22}
$$
$$
u = i-R, i-R+1, \ldots, i+R
$$

Replacing (22) into (21), we get

$$
\begin{aligned}
e_i(w_k) - e_i(w_{k-1}) &\leq -\eta(\Delta w_{k-1})^T (\Delta w_{k-1})(1-\frac{1}{2}\eta z) \\
&= -\eta(1-\frac{1}{2}\eta z)\sum_{u=i-R}^{i+R}[d(i)-x_k(i)]^2 f'^2(s_{k-1}(i)) x_{k-1}^2(u) \\
&= -\eta[d(i)-x_k(i)]^2 (1-\frac{1}{2}\eta z) f'^2(s_{k-1}(i))\sum_{u=i-R}^{i+R} x_{k-1}^2(u) \\
&= -e_i(w_{k-1})\eta(2-\eta z) f'^2(s_{k-1}(i))\sum_{u=i-R}^{i+R} x_{k-1}^2(u)]
\end{aligned}
$$
(23)

Set $a_{k-1}(i) = f'^2(s_{k-1}(i))\sum_{u=i-R}^{i+R} x_{k-1}^2(u)$, $i=1,2,\cdots,m$, and the minimum of these $m$ value is $a_{(k-1)\min}$, so

$$
\begin{aligned}
E_{k+1} &= \sum_{i=1}^{m} e_{k+1}(i) = \sum_{i=1}^{m} e_i(w_k) \\
&= \sum_{i=1}^{m} e_i(w_{k-1})[1-\eta(2-\eta z) a_{k-1}(i)] \\
&\leq [1-\eta(2-\eta z) a_{(k-1)\min}]\sum_{i=1}^{m} e_i(w_{k-1}) \\
&= [1-\eta(2-\eta z) a_{(k-1)\min}]\sum_{i=1}^{m} e_k(i) \\
&= [1-\eta(2-\eta z) a_{(k-1)\min}] E_k
\end{aligned}
$$
(24)

Therefore,

$$
\begin{aligned}
E_{k+1} &\leq [1-\eta(2-\eta z) a_{(k-1)\min}] E_k \\
&\leq [1-\eta(2-\eta z) a_{(k-1)\min}][1-\eta(2-\eta z) a_{(k-2)\min}] E(w_{k-1}) \\
&\leq [1-\eta(2-\eta z) a_{(k-1)\min}][1-\eta(2-\eta z) a_{(k-2)\min}] L [1-\eta(2-\eta z) a_{1\min}] E_2
\end{aligned}
$$
(25)

The minimum of $a_{1\min}, a_{2\min}, \cdots, a_{(k-1)\min}$ is denoted as $a_{\min}$, then (23) becomes

$$E_{k+1} \le [1 - \eta(2 - \eta z)a_{\min}]^{k-1} E_2 = g^{k-1}(\eta) E_2 \qquad (26)$$

where

$$g(\eta) = 1 - \eta(2 - \eta z)a_{\min} \qquad (27)$$

As $E_k \ge 0$, when $0 < g(\eta) < 1$, $E$ converges to zero exponentially as $k \to \infty$.

Therefore, the weights will not change and converge to $w_c(i, u)$. It means this algorithm has the ability to find the transition rules with correct weights $w_c(i, u)$.

### 3.3 The convergence conditions of CAW

We discuss the detailed conditions of $\nabla^2 e_i(w)^T$ this section. As we know,

$$e_i(w)\big|_{w=\varepsilon_{k-1}} = \frac{1}{2}[d(i) - f(\sum_{u=i-R}^{i+R} x_{k-1}(u))w(i,u))]^2\big|_{w=\varepsilon_{k-1}}$$
$$= \frac{1}{2}[d(i) - f(\sum_{u=i-R}^{i+R} x_{k-1}(u))\varepsilon_{k-1}(i,u))]^2, i = 1,2,L,m \qquad (28)$$

From (7) and (13), we get

$$\frac{\partial^2 e_i(w)}{\partial w(i,u)\partial w(i,v)}\bigg|_{w=\varepsilon_{k-1}} = -\frac{\partial[d(i) - x_k(i)]f'(s_{k-1}(i))x_{k-1}(u)}{\partial w(i,v)}\bigg|_{w=\varepsilon_{k-1}}$$
$$= f'(s_{k-1}(i))x_{k-1}(v)f'(s_{k-1}(i))x_{k-1}(u)$$
$$-[d(i) - x_k(i)]f''(s_{k-1}(i))x_{k-1}(v)x_{k-1}(u)$$
$$= x_{k-1}(v)x_{k-1}(u)\{(f'(s_{k-1}(i)))^2$$
$$-[d(i) - x_k(i)]f''(s_{k-1}(i))\} \qquad (29)$$

which is independent of $\varepsilon_{k-1}$.

Therefore,

$$\nabla^2 e_i(w)^T = \{(f'(s_{k-1}(i)))^2 - [d(i) - x_k(i)]f''(s_{k-1}(i))\}$$
$$\begin{bmatrix} x_{k-1}^2(i-R) & x_{k-1}(i-R)x_{k-1}(i-R+1)\cdots x_{k-1}(i-R)x_{k-1}(i+R) \\ x_{k-1}(i-R+1)x_{k-1}(i-R) & x_{k-1}^2(i-R+1) & \cdots x_{k-1}(i-R+1)x_{k-1}(i+R) \\ \vdots & \vdots & \vdots \\ x_{k-1}(i+R)x_{k-1}(i-R) & x_{k-1}(i+R)x_{k-1}(i-R+1)\cdots & x_{k-1}^2(i+R) \end{bmatrix}$$
$$= H_{k-1}(i)X_{k-1}(i) \qquad (30)$$

where

$$H_{k-1}(i) = \{(f'(s_{k-1}(i)))^2 - [d(i) - x_k(i)]f''(s_{k-1}(i))\} \qquad (31)$$

$$X_{k-1}(i) = \begin{bmatrix} x_{k-1}^2(i-R) & x_{k-1}(i-R)x_{k-1}(i-R+1)\dots x_{k-1}(i-R)x_{k-1}(i+R) \\ x_{k-1}(i-R+1)x_{k-1}(i-R) & x_{k-1}^2(i-R+1) & \dots x_{k-1}(i-R+1)x_{k-1}(i+R) \\ \vdots & \vdots & \vdots \\ x_{k-1}(i+R)x_{k-1}(i-R) & x_{k-1}(i+R)x_{k-1}(i-R+1)\dots & x_{k-1}^2(i+R) \end{bmatrix} \qquad (32)$$

where $X$ is a positive semidefinite symmetric matrix, and it has $2R+1$ eigenvalues, in which there is only one nonzero eigenvalue

$$\lambda_{X_{k-1}(i)} = \sum_{u=i-R}^{i+R} x_{k-1}^2(i,u)(\lambda_X \ge 0) \qquad (33)$$

If $H_{k-1}(i) \le 0$, $\nabla^2 e_i(w)^T$ is negative semidefinite, its maximum eigenvalue is zero, (27) becomes $g(\eta) = 1 - 2\eta a_{\min}$. Therefore, $E$ converges to zero exponentially as $k \to \infty$ under the condition

$$0 < 2\eta a_{\min} < 1 \qquad (34)$$

As $g'(\eta) = -2a_{\min} < 0$, the convergence speed increases with $\eta$.

If $H_{k-1}(i) > 0$, $\nabla^2 e_i(w)^T$ is positive semidefinite, and its norm is

$$C = \left\|\nabla^2 e_i(w)^T\right\| = \left\|H_{k-1}(i)X_{k-1}(i)\right\| = H_{k-1}(i)\lambda_{X_{k-1}(i)} \qquad (35)$$

Because $0 < g(\eta) < 1$, $g'(\eta) = 2a_{\min}(\eta C - 1)$ we get

$$0 \le C \le \frac{2}{\eta} \text{ when } 2\eta a_{\min} \le 1 \ (\ g'(\eta) \text{ depends on } \eta C\ ) \qquad (36)$$

or

$$\frac{2}{\eta}(1 - \frac{1}{2\eta a_{\min}}) < C \le \frac{2}{\eta} \text{ when } 2\eta a_{\min} > 1 \ (\ g'(\eta) \text{ depends on } \eta C\ ) \qquad (37)$$

It is easy to find $f$ and $\eta$ satisfying the above condition in the real field, such as Sigmoid function and linear bounded functions. Note that these are sufficient conditions for the convergence of $E$, not necessary conditions.

## 4. Simulation experiments and its analysis

### 4.1 The CAW with correct rules

Take one-dimensional CAW with radius $r=1$, the number of cells $N=10$ and learning factor $\eta = 3$ as an example. Initial state values, expected states and initial weights are generated randomly. The transition function is Sigmoid function $f = \frac{1}{1+e^{-x}}$. The evolution configuration and error curve of one-dimensional CAW with fixed boundary are depicted in Figure 3, where red dots represent cells at corresponding spots and the volumes of dots increase with the states of cells. The blue dots in the last column are desired fixed states.

In Figure 3(a), the output before the 50th step is the process of finding rules. After getting $w_c(i,u)$, which satisfies $T(d(i)) = d(i)$, the output of CAW with correct rules from an arbitrary initial states is shown from step 50 to 100. Each cell arrives at its desired states and the whole CAW relax to the fixed configuration.

**Figure 3** The evolution configuration and error curve of one-dimensional CAW with $\eta = 3$

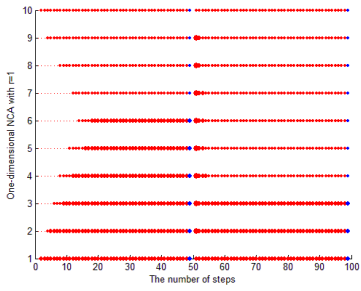

(a) The evolution configuration (NCA)

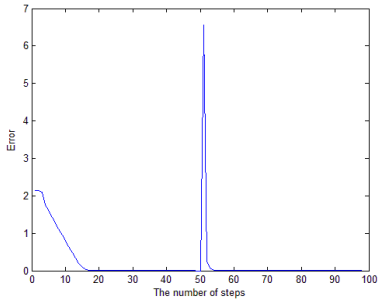

(b) The error curve

### 4.2 *The convergence speed experiment of CAW*

Take one-dimensional CAW with $r=3$, $N=50$, $\eta = 2$, initial states 0, expected states 1 and initial weights 0.1 as an example. The transition function is Sigmond function, too. The evolution configuration and error curve of one-dimensional CAW with fixed boundary finding rules are shown in Figure 4. The error is 0.8873 at the 5th step, and 0.0169 at the 50th step, which converges rapidly.

**Figure 4** The evolution configuration and error curve of one-dimensional CAW with $\eta = 2$

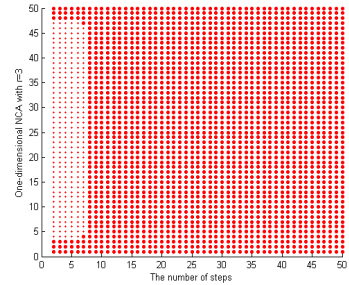

(a) The evolution configuration (NCA)

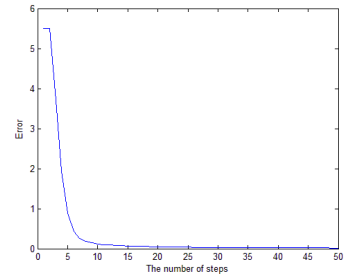

(b) The error curve

**Table 2.** The error of CAW with some learning factors and radius at the 200th iteration.

| radius $\eta$ | $r$=1 | $r$=2 | $r$=3 | $r$=4 | $r$=5 |
|---|---|---|---|---|---|
| 0.1 | 0.3829 | 0.1710 | 0.1037 | 0.0718 | 0.0533 |
| 1 | 0.0221 | 0.0122 | 0.0082 | 0.0060 | 0.0046 |
| 2 | 0.0105 | 0.0059 | 0.0040 | 0.0029 | 0.0023 |
| 4 | 0.0051 | 0.0029 | 0.0019 | 0.0014 | 0.0011 |
| 8 | 0.0025 | 0.0014 | 0.0009 | 0.0006 | 0.0004 |
| 16 | 0.0012 | 0.0004 | 0.0001 | 0 | 0 |

**Table 3.** The error, $C$ and $a_{\min}$ of CAW with some learning factors and radius at the 200th iteration.

| radius $\eta$ | $r$=1 | $r$=3 | $r$=10 | $r$=20 |
|---|---|---|---|---|
| 0.1 | 0.0047 ($C = 0.1337$, $a_{\min} = 0.0019$) | 0.0013 ($C = 0.4763$, $a_{\min} = 0.0044$) | 0 ($C = 4.5765$, $a_{\min} = 0.0128$) | 0 ($C = 8.8153$, $a_{\min} = 0.0236$) |
| 2 | 0 | 0 | 0 | 0 |
|   | ($C = 0.1301$ $a_{\min} = 0.0019$ | ($C = 0.4200$ $a_{\min} = 0.0044$) | ($C = 2.3928$ $a_{\min} = 0.0128$) | ($C = 6.919$ $a_{\min} = 0.0236$) |
| 8 | 0 ($C = 0.1237$ $a_{\min} = 0.0019$) | 0 ($C = 0.3837$ $a_{\min} = 0.0044$) | convergent with oscillation ($C = 3.5202$ $a_{\min} = 0.0079$) | divergent ($C = 11.0987$ $a_{\min} = 0$) |
| 16 | 0 ($C = 0.1232$ $a_{\min} = 0.0019$) | 0 ($C = 0.3781$ $a_{\min} = 0.0044$) | divergent ($C = 7.6694$ $a_{\min} = 0$) | divergent ($C = 12.7492$ $a_{\min} = 0$) |
| 32 | 0 ($C = 0.1224$ $a_{\min} = 0.0019$) | divergent ($C = 2.8698$ $a_{\min} = 0$) | divergent ($C = 8.9998$ $a_{\min} = 0$) | divergent ($C = 12.75$ $a_{\min} = 0$) |

Table 2 indicates the error of CAW with some learning factors and radiuses at the 200th step. It is obvious that the convergence speed increases as learning factor increases for CAW with the same radius, and convergence speed increases as radius increases for CAW with the same learning factor, which verifies (34) under the condition $H_{k-1}(i) \le 0$ with $d(i) = 1$ and $a_{\min} \to 0$ (Because the derivative of sigmoid function is zero as its output $x(i) \to d(i)$).

*4.3 The convergence conditions experiment of CAW*

To verify (36) and (37), we choose $d(i) = 0.5$, initial states 0.1 and $H_{k-1}(i) > 0$. The error of CAW with some learning factors and radiuses at the 200th step is shown in Table 3.

It can be seen that

(1) If $2\eta a_{\min} \le 1$ and $0 \le C \le \dfrac{2}{\eta}$, $E$ converges, such as CAW with $r=1$, $\eta = 0.1, 2, 8, 16$ and so on. Note that when $r=1$ and $\eta = 32$, $C = 0.1224 > \dfrac{2}{32}$; when $r=3$ and $\eta = 8$, $C = 0.3837 > \dfrac{2}{8}$. The above conditions do not satisfy (36), because (36) is sufficient condition, not necessary.

(2) In fact, (37) doesn't appear in the example, because the derivative of Sigmoid function is zero when it comes into saturation, which makes $a_{\min} = 0$.

(3) If $E$ is convergent with oscillation, $a_{\min}$ is the same for CAW with the same radius.

## 5. Conclusion

In this paper, We present CA with weights(CAW), which have ample physical meanings and extend the category of CA. Then as for CAs which relax to a fixed configuration, a method to find transition rules with an exponential speed is discussed. The CAW with the correct rules will relax to the fixed configuration regardless of the initial configuration.

## 6. Acknowledgments

This work is supported by International Collaborative Research Project of Qilu University of Technology (No.QLUTGJHZ2018020), Youth Innovation Science and technology support plan of colleges in Shandong Province (No.2021KJ025), National Nature Science Foundation of China (No.61903207), Basic research (free-exploration) project of Shenzhen Science and Innovation Commission (201803023000889), Major scientic and technological innovation projects of Shandong Province (No.2019JZZY010731 and No.2020CXGC010901), and this research received no external funding.

## References

[1] J. von Neumann, "Theory of Self-Reproducing Automata", A. W. Burks University of Illinois Press, Urbana, ILL, USA, 1966.

[2] C. B. Jiang, R. L. Li, T. G. Chen, C. H. Xu, L. Li and S. F. Li, "A two-lane mixed traffic flow model with drivers' intention to change lane based on cellular automata," International Journal of Bio-Inspired Computation, vol. 16, no. 4, pp. 229-240, 2020, doi: 10.1504/IJBIC.2020.10034141.

[3] Gheorghe, Adrian V. , and D. V. Vamanu . "Forest fire essentials: a cellular automaton-wise, percolation-oriented model." International Journal of Critical Infrastructures 4.4(2008):430-444, doi: 10.1504/IJCIS.2008.020161.

[4] Zhou, G. , Q. Wu , and A. Chen . "Research of cellular automata model for forest fire spreading simulation." Yi Qi Yi Biao Xue Bao/Chinese Journal of entific Instrument 38.2(2017):288-294, doi: 10.19650/j.cnki.cjsi.2017.02.004.

[5] Y. Piao, X. Li, M. Zhang, J. Yu and H. Lu, "Saliency Detection via Depth-Induced Cellular Automata on Light Field," in IEEE Transactions on Image Processing, vol. 29, pp. 1879-1889, 2020, doi: 10.1109/TIP.2019.2942434.

[6] M. Trevisi, A. Akbari, M. Trocan, Á. Rodríguez-Vázquez and R. Carmona-Galán, "Compressive

Imaging Using RIP-Compliant CMOS Imager Architecture and Landweber Reconstruction," in IEEE Transactions on Circuits and Systems for Video Technology, vol. 30, no. 2, pp. 387-399, Feb. 2020, doi: 10.1109/TCSVT.2019.2892178.

[7]  A. Sanchez-Macian, A. Martin-Toledano, J. A. Bravo-Montes, F. Garcia-Herrero and J. A. Maestro, "Reducing the Impact of Defects in Quantum-Dot Cellular Automata (QCA) Approximate Adders at Nano Scale," in IEEE Transactions on Emerging Topics in Computing, doi: 10.1109/TETC.2021.3136204.

[8]  Zhang, N. X. , Zhu, H. B., Lin, H., et al. "One-dimensional cellular automaton model of traffic flow considering dynamic headway." Acta Physica Sinica 64.2(2015), doi: 10.7498/aps.64.024501.

[9]  F. Enayatollahi, M. A. A. Atashgah, S. M. -B. Malaek and P. Thulasiraman, "PBN-Based Time-Optimal Terminal Air Traffic Control Using Cellular Automata," in IEEE Transactions on Aerospace and Electronic Systems, vol. 57, no. 3, pp. 1513-1523, June 2021, doi: 10.1109/TAES.2020.3048787.

[10] Lv, W. , Wang, J. H., Fang, Z. M., et al. " Simulation method of urban evacuation based on mesoscopic cellular automata." Acta Physica Sinica 70.10(2021), doi: 10.7498/aps.70.20210018.

[11] Fangfang Zhang, Junfei Qiao, Chaobin Liu, Naigong Yu and Xiaogang Ruan, "A Cellular Automata Model for the Sequencing Batch Reactor of Activated Sludge Processes," 2006 6th World Congress on Intelligent Control and Automation, 2006, pp. 1558-1562, doi: 10.1109/WCICA.2006.1712612.

[12] Qiao, J. , and L. I. Rong . "The Simulation for Treatment Process of Activated Sludge with Three-Dimensional Cellular Automata." Information and Control (2010), doi: 10.3969/j.issn.1002-0411.2010.01.017.

[13] M. -A. I. Tsompanas, G. C. Sirakoulis and A. I. Adamatzky, "Evolving Transport Networks With Cellular Automata Models Inspired by Slime Mould," in IEEE Transactions on Cybernetics, vol. 45, no. 9, pp. 1887-1899, Sept. 2015, doi: 10.1109/TCYB.2014.2361731.

[14] N. Shimada and H. Torikai, "A Novel Asynchronous Cellular Automaton Multicompartment Neuron Model," in IEEE Transactions on Circuits and Systems II: Express Briefs, vol. 62, no. 8, pp. 776-780, Aug. 2015, doi: 10.1109/TCSII.2015.2433471.

[15] T. Matsubara and H. Torikai, "An Asynchronous Recurrent Network of Cellular Automaton-Based

Neurons and Its Reproduction of Spiking Neural Network Activities," in IEEE Transactions on Neural Networks and Learning Systems, vol. 27, no. 4, pp. 836-852, April 2016, doi: 10.1109/TNNLS.2015.2425893.

[16] L. Sallemi, I. Njeh and S. Lehericy, "Towards a Computer Aided Prognosis for Brain Glioblastomas Tumor Growth Estimation," in IEEE Transactions on NanoBioscience, vol. 14, no. 7, pp. 727-733, Oct. 2015, doi: 10.1109/TNB.2015.2450365.

[17] K. Takeda and H. Torikai, "A Novel Hardware-Efficient Cochlea Model Based on Asynchronous Cellular Automaton Dynamics: Theoretical Analysis and FPGA Implementation," in IEEE Transactions on Circuits and Systems II: Express Briefs, vol. 64, no. 9, pp. 1107-1111, Sept. 2017, doi: 10.1109/TCSII.2017.2672824.

[18] Zhang, X. L., Tan, H. L., Tang, G. N., et al. " A cellular automaton model for electrocardiogram considering the structure of heart." Acta Physica Sinica 66.20(2017), doi: 10.7498/aps.66.200501.

[19] Stephen Wolfram. Universality and complexity in cellular automata, 1984, 10(1-2), 1-35. doi:10.1016/0167-2789(84)90245-8.

[20] Cui Zh., Zhang J., Wang Y., Cao Y., et al. " A pigeon-inspired optimization algorithm for many-objective optimization problems." SCIENCE CHINA Information Sciences, 2019, 62(7): 070212

[21] Zhihua Cui, Lihong Zhao, Youqian Zeng, Yeqing Ren, Wensheng Zhang, Xiao-Zhi Gao, A Novel PIO Algorithm with multiple selection strategies for many-objective optimization problems, Complex System Modeling and Simulation, 2021,4(1): 291-307

[22] Xingjuan Cai, Shaojin Geng, Di Wu, Jinjun Chen, Unified integration of many-objective optimization algorithm based on temporary offspring for software defects prediction. Swarm and Evolutionary Computation, 2021, 63:100871.